\documentclass[twocolumn,english,aps,pra,longbibliography, superscriptaddress, floatfix]{revtex4-2}
\usepackage[utf8]{inputenc}
\usepackage{amsmath}
\usepackage{amssymb}
\usepackage{braket}
\usepackage{color}
\usepackage{graphicx}
\usepackage{pifont}
\usepackage{xcolor}
\usepackage[colorlinks=true,urlcolor=blue,citecolor=blue,linkcolor=blue]{hyperref}
\usepackage{setspace}
\usepackage{siunitx}

\begin{document}

\title{Pulse variational quantum eigensolver on cross-resonance-based hardware}

\author{Daniel J. Egger}
\email{deg@zurich.ibm.com}
\affiliation{IBM Quantum -- IBM Research Europe - Zurich, S\"aumerstrasse 4, 8803 R\"uschlikon, Switzerland}
\author{Chiara Capecci}
\affiliation{Dipartimento di Ingegneria e Scienze dell'Informazione e Matematica, Universit\`a degli Studi dell'Aquila, 67100 Coppito, L'Aquila, Italy}
\author{Bibek Pokharel}
\affiliation{IBM Quantum -- IBM Research Almaden, 650 Harry Road, San Jose, CA 95120, USA}
\author{Panagiotis Kl. Barkoutsos}
\altaffiliation{Presently at PASQAL}
\affiliation{IBM Quantum -- IBM Research Europe - Zurich, S\"aumerstrasse 4, 8803 R\"uschlikon, Switzerland}
\author{Laurin E. Fischer}
\affiliation{IBM Quantum -- IBM Research Europe - Zurich, S\"aumerstrasse 4, 8803 R\"uschlikon, Switzerland}
\affiliation{Theory and Simulation of Materials (THEOS), {\'E}cole Polytechnique F{\'e}d{\'e}rale de Lausanne, 1015 Lausanne, Switzerland}
\author{Leonardo Guidoni}
\affiliation{Dipartimento di Scienze Fisiche e Chimiche, Universit\`a degli Studi dell'Aquila, 67100 Coppito, L'Aquila, Italy}
\author{Ivano Tavernelli}
\affiliation{IBM Quantum -- IBM Research Europe - Zurich, S\"aumerstrasse 4, 8803 R\"uschlikon, Switzerland}
\date{\today}

\begin{abstract}
State-of-the-art noisy digital quantum computers can only execute short-depth quantum circuits.
Variational algorithms are a promising route to unlock the potential of noisy quantum computers since 
the depth of the corresponding circuits can be kept well below hardware-imposed limits.
Typically, the variational parameters correspond to virtual $R_Z$ gate angles, implemented by phase changes of calibrated pulses.
By encoding the variational parameters directly as hardware pulse amplitudes and durations we succeed in further shortening the pulse schedule and overall circuit duration.
This decreases the impact of qubit decoherence and gate noise. 
As a demonstration, we apply our pulse-based variational algorithm to the calculation of the ground state of different hydrogen-based systems (H$_2$, H$_3$ and H$_4$) using IBM cross-resonance-based hardware.
We observe a reduction in schedule duration of up to $5\times$ compared to CNOT-based Ans\"atze, while also reducing the measured energy. 
In particular, we observe a sizable improvement of the minimal energy configuration of H$_3$ compared to a CNOT-based variational form.
Finally, we discuss possible future developments including error mitigation schemes and schedule optimizations, which will enable  further improvements of our approach paving the way towards the simulation of larger systems on noisy quantum devices.
\end{abstract}

\maketitle

\section{Introduction}

Current quantum computers are noisy and are constituted of qubits with finite coherence times.
This bounds the depth of the circuits that they can reliably execute.
There is thus a large interest 
in short-depth noise-resilient algorithms such as the variational quantum algorithm (VQA)~\cite{Moll2018}. 
VQAs can be applied to quantum chemistry~\cite{OMalley2016, McClean2016, ollitrault2022quantum, motta2021emerging, OBrien2022}, machine learning~\cite{Havlicek2019, Abbas2021, Melo2022} and optimization~\cite{Farhi2014} tasks.
In a VQA, the expectation value $\braket{\psi(\boldsymbol{\theta})|\mathcal{O}|\psi(\boldsymbol{\theta})}$ of an observable $\mathcal{O}$ is optimized by varying the parameters $\boldsymbol{\theta}$ of a trial variational state $\ket{\psi(\boldsymbol{\theta})}$.
Typically, the variational state is prepared by a parameterized quantum circuit, the Ansatz.
For example, in quantum chemistry $\ket{\psi(\boldsymbol{\theta})}$ can be prepared with the unitary coupled cluster with singles and doubles (UCCSD) Ansatz~\cite{Barkoutsos2018}.
For combinatorial optimization, the quantum approximate optimization algorithm prepares a trial state by alternating applications of a cost-function operator and a mixer operator~\cite{Harrigan2021, Santra2022, Pelofske2023, Sack2023}.
However, many Ans\"atze are often still too deep for execution on current quantum hardware~\cite{Barkoutsos2018, Weidenfeller2022}.
This has spurred an interest to generate variational states with more ressource-efficient circuits~\cite{Kandala2017, Ryabinkin2018, Grimsley2019, Lee2019, Egger2021}.
In superconducting qubits~\cite{Krantz2019, Kjaergaard2020, Koch2007}, the circuit instructions are translated into micro-wave pulses that manipulate the quantum information.
For example, on IBM Quantum systems all circuits are broken down into the hardware-native basis gates $\{X, \sqrt{X},{\rm CNOT}, R_Z(\theta)\}$.
Here, the $X$, $\sqrt{X}$, and $\rm CNOT$ gates are implemented by carefully calibrated pulses.
All the parameters of a circuit are therefore encoded in the virtual-$Z$ rotations $R_Z(\theta)$, i.e. zero-duration instructions that only change the phase of subsequent pulses~\cite{McKay2017}.
The duration of the Ansatz is thus independent of the optimization parameters $\boldsymbol{\theta}$.

The pulses implementing the native basis gate set are carefully calibrated \textit{a priori}, a costly task typically done with error amplifying gate sequences~\cite{Tornow2022} and sometimes optimal control~\cite{Werninghaus2020}.
Quantum optimal control (OC) has a long history~\cite{Glaser2015, Boscain2021, Koch2022} and provides methods to create quantum states~\cite{Bao2018, Malis2019}, gates~\cite{Egger2013, Kelly2014, Winick2021}, and control non-unitary dynamics~\cite{Koch2016} such as a measurement process~\cite{Egger2014b, Boutin2017}.
However, in superconducting qubits, model inaccuracies make it difficult to apply pulses generated through simulations~\cite{Egger2014}.
One must either improve the model~\cite{Wittler2021} or resort to closed-loop optimal control on the hardware~\cite{Egger2014}.
In closed-loop optimal control, a cost function, which can correspond to a gate fidelity, is optimized by an algorithm that varies parameters in a parameterized pulse shape~\cite{Kelly2014, Werninghaus2020}.
Similarly, in a VQA the expectation value of an observable is optimized by varying parameters in a parameterized quantum circuit which is ultimately lowered to pulses.
Closed-loop optimal control and VQAs can thus be viewed as the same task~\cite{Magann2021}.

Currently, VQAs with the basis gate set $\{X, \sqrt{X},$ ${\rm CNOT}, R_Z(\theta)\}$ amount to optimizing phases while OC optimizes pulse parameters.
Pulse-level control of cloud-based quantum computers~\cite{Mckay2018, Alexander2020} enables a direct optimization of the pulse parameters in the variational quantum eigensolver (VQE)~\cite{Peruzzo2014}.
This has been explored in previous works, which we briefly summarize.
The authors of Ref.~\cite{Liang2022b} show how to optimize only the amplitude of pulses on cross-resonance (CR) systems~\cite{Sheldon2016} to increase the accuracy of a binary classification.
On the other hand, using the PAN ansatz~\cite{Liang2022} the authors let the variational algorithm change the amplitude and frequency of the pulses.
However, optimizing only pulse amplitudes and not durations makes it impossible to mitigate decoherence by adaptively shortening the pulse schedule.
Indeed, this is what has been proposed in Ctrl-VQE where both the duration and the amplitude of square pulses is optimized~\cite{Meitei2021}.
Numerical simulations show that leakage outside of the computational space can reduce the state preparation time in Ctrl-VQE and improve results~\cite{Asthana2022}.
Finally, in Ref.~\cite{Meirom2022} the authors study a pulse-based variational Ansatz in which the duration of two-qubit CR gates is optimized at a fixed amplitude.
They show numerical simulations that achieve chemical accuracy on molecules that require up to four qubits and present hardware results for H$_2$ on two qubits.

In this work, we simultaneously optimize single-qubit pulses as well as both the duration and the amplitude of CR pulses.
Here we demonstrate a pulse-based VQE that optimizes both duration and amplitude with systems that use up to eight qubits.
Furthermore, we perform the full VQE parameter optimization on quantum hardware.
This allows the optimizer to capture $T_1$ and $T_2$ related tradeoffs that favor short and intense pulses.
For context, the largest VQE by qubit count on hardware was done on 20 qubits with parameters optimized in a noiseless simulation~\cite{Google2023}.
In addition, we discuss the implementation of advanced error mitigation in pulse-based VQAs which is tricky since the effect of the pulses is hard to capture.

In Sec.~\ref{sec:cr_dynamics} we review the dynamics of the CR gate and existing connections to variational algorithms.
In Sec.~\ref{sec:chem} we introduce the chemical systems that we study, namely, H$_2$, H$_3$, and H$_4$.
Next, in Sec.~\ref{sec:hardware_results}, we study these systems on devices with up to eight qubits.
We discuss error mitigation in Sec.~\ref{sec:error_mit} and conclude in Sec.~\ref{sec:conclusion}.

\section{Cross-resonance dynamics and variational quantum algorithms\label{sec:cr_dynamics}}

\begin{figure}[!]
    \centering
    \includegraphics[width=\columnwidth]{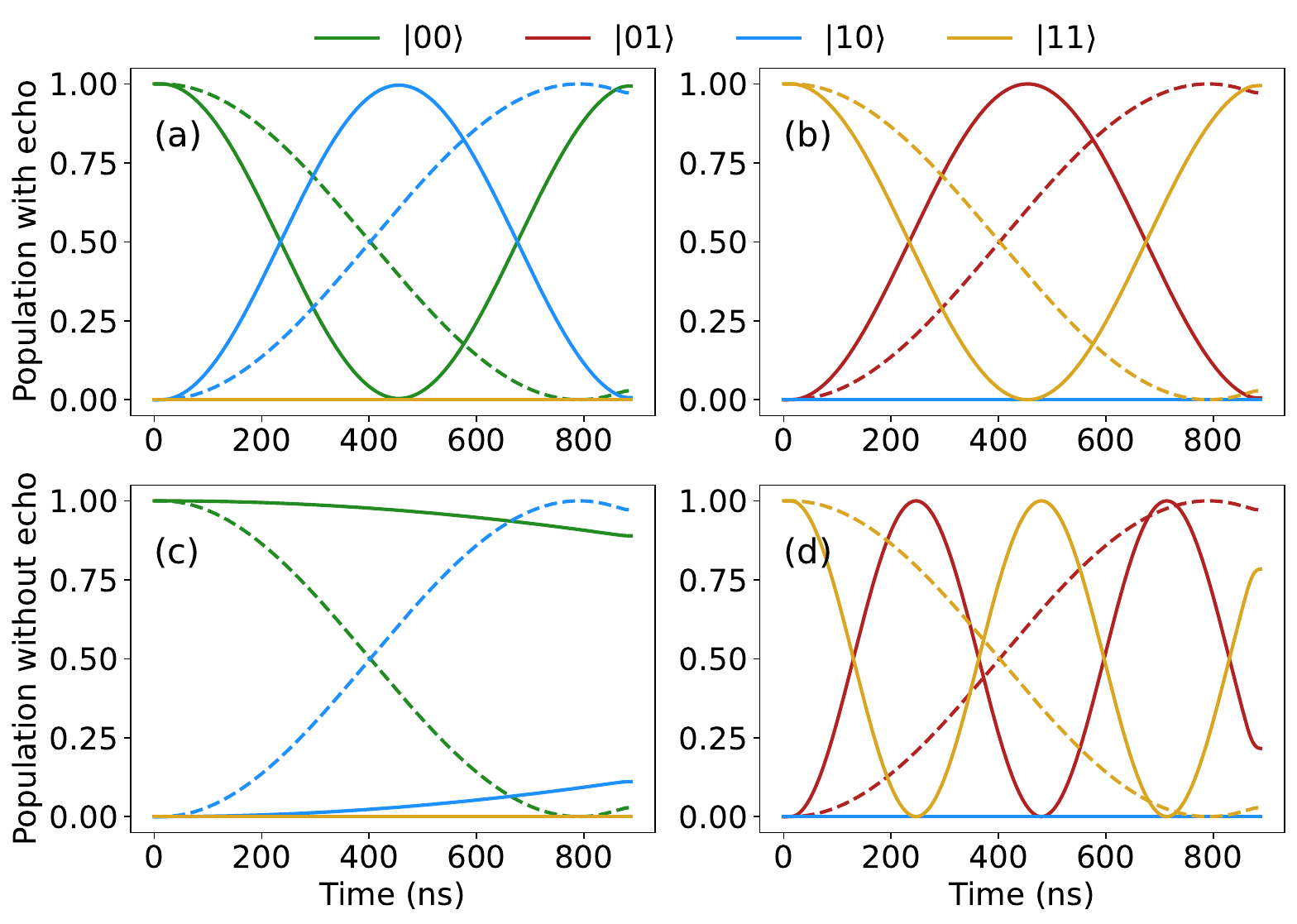}
    \caption{
    \textbf{Dynamics of the cross-resonance gate.}
    The states are labeled according to $\ket{\text{target},\text{control}}$.
    The simulation is done (a), (b) with an echo and (c), (d) without an echo.
    In both cases, the strength of the $\omega_{ij}$ in Eq.~(\ref{eq:cr_eff}) is measured on qubits (0, 1) of \emph{ibm\_lagos} with Hamiltonian tomography~\cite{QiskitExperiments}.
    The strength of the cross-resonance terms without an echo is reported in Appendix~\ref{sec:hardware}.
    With an echo we measure $\omega_{ZX}=872(2)~{\rm kHz}$, $\omega_{ZY}=715(2)~{\rm kHz}$, $\omega_{ZZ}=-35(1)~{\rm kHz}$, $\omega_{IX}=53(1)~{\rm kHz}$, $\omega_{IY}=-69(2)~{\rm kHz}$, and $\omega_{IZ}=-35(1)~{\rm kHz}$.
    The dashed line shows the population when only the $ZX$ term is retained.
    }
    \label{fig:cr_dynamics}
\end{figure}

Dispersively coupled fixed-frequency transmon qubits can be entangled with the CR interaction~\cite{Rigetti2010, Sheldon2016, Fischer2022b}.
Here, one qubit, the control, is driven at the frequency of the other, the target.
The resulting effective Hamiltonian is
\begin{align}\label{eq:cr_eff}
    \Bar{H}_{cr}=\frac{1}{2}\left(Z\otimes B + I\otimes C\right)
\end{align}
where $B=\omega_{ZI}I+\omega_{ZX}X+\omega_{ZY}Y+\omega_{ZZ}Z$ and $C=\omega_{IX}X+\omega_{IY}Y+\omega_{IZ}Z$.
Here, $X$, $Y$, and $Z$ are Pauli matrices and $I$ is the identity. 
The coefficients $\omega_{ij}$ are the strength of the CR interaction.
They depend on the properties of the qubits and the drive strength~\cite{Magesan2020}.
To illustrate the dynamics of the cross-resonance gate we first measure the $\omega_{ij}$'s with Hamiltonian tomography~\cite{Sheldon2016} implemented in Qiskit Experiments.
Next, we simulate Eq.~(\ref{eq:cr_eff}) with Qiskit Dynamics.
The strongest terms in $\Bar{H}_{cr}$ are $\omega_{ZX}$ and $\omega_{IX}$, see Fig.~\ref{fig:cr_dynamics}.
The CR interaction is typically used to engineer a $\rm CNOT$ gate by eliminating the non-$ZX$ terms with an echo sequence and cancellation tones~\cite{Sundaresan2020}.
The phase of the CR drive controls the relative magnitude of the $\omega_{ZX}$ and $\omega_{ZY}$ coefficients.
Simulations of the time-evolution under $\bar{H}_{cr}$ and Hamiltonian tomography show that the echo cancels the large $\omega_{IX}$ term, see Fig.~\ref{fig:cr_dynamics}.
With or without an echo, the resulting entanglement is usable in VQAs by replacing $\rm CNOT$ gates by CR tones with fixed parameters, as done in Ref.~\cite{Ibrahim2022}.
Here, the resulting Ans\"atze are parameterized by virtual-$Z$ rotations.

The $\omega_{ij}$'s are non-linear with drive amplitude~\cite{Magesan2020} but the rotation implemented by $\exp(-i \tau \Bar{H}_{cr})$ is linear in time $\tau$.
The linearity in $\tau$ enables transpiler passes to create $R_{ZX}(\theta)$ rotations built from calibrated CNOT gates by scaling the duration of cross-resonance pulses~\cite{Earnest2021}. 
The resulting shorter pulses reduce hardware errors in quantum approximate optimization~\cite{Earnest2021, Weidenfeller2022} and machine learning~\cite{Melo2022} which are both a form of pulse-based VQA.

\section{Test Systems: Applications to Quantum Chemistry\label{sec:chem}}

\begin{figure}[h!]
    \centering
    \includegraphics[width=\columnwidth]{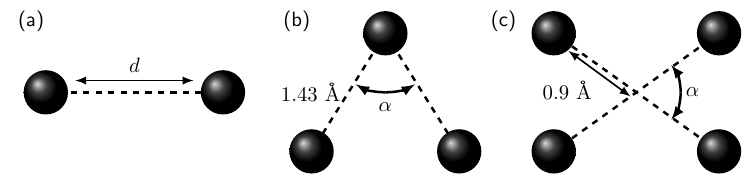}
    \caption{\textbf{Considered hydrogen-based systems.} (a) For H$_2$ we vary the distance $d$.
    For H$_3$ (b) we vary the angle~$\alpha$ and for H$_4$ (c) we consider the angle $\alpha=40^\circ$.}
    \label{fig:molecules}
\end{figure}

In this work, we investigate pulse-based VQE on hydrogen-based systems. 
We consider molecular hydrogen H$_2$, the triangular H$_3$ and the rectangular H$_4$~\cite{Greene2021}, see Fig~\ref{fig:molecules}. 
The triangular H$_3$ is highly frustrated and the rectangular H$_4$ has strong correlations making both systems interesting to study. 
We model all the systems in the minimal Gaussian basis set Slater-type orbital (STO)-3G.
The fermionic problem is mapped to a Hamiltonian expressed as the linear combination $H=\sum_i\alpha_iP_i$.
Here, the $\alpha_i$'s are coefficients and the $P_i$'s are Pauli operators made of tensor products of single-qubit Pauli matrices $I$, $X$, $Y$, and $Z$, e.g., $YZYXXI$.
A parameterized circuit Ansatz creates a state $\ket{\psi(\boldsymbol{\theta})}$ on the quantum hardware.
As a cost function we minimize the energy, i.e., ${\rm min}_{\boldsymbol{\theta}}\langle\psi(\boldsymbol{\theta})|H|\psi(\boldsymbol{\theta})\rangle$, in a closed-loop with the hardware.
To reduce the number of quantum circuits to measure we group the Pauli operators $\{P_i\}$ into qubit-wise commuting groups.
This allows us to measure each group with a single basis-change before the final measurement.

In the STO-3G basis each hydrogen atom requires two qubits to model, one for each spin orbital.
H$_2$, the exception, is mapped to spin operators with the parity mapping~\cite{Seeley2012} and a reduction to two-qubits, leveraging particle number conservation, resulting in five Pauli terms.
The H$_2$ dissociation curve is often studied as a benchmark for VQAs~\cite{OMalley2016, Kandala2017, Meirom2022, Ratini2022}.
H$_3$ has a triangular geometry, and we map this fermionic system to six qubits with the Jordan-Wigner transformation. 
The resulting 62 Pauli terms are measured in 21 sets of qubit-wise commuting elements.
Initially, we compute the dissociation curve for the equilateral system using the classical full configuration interaction (full CI) method to find the bond distance. 
We then study the dissociation curve of a more general isosceles conformation with two equal sides of the triangle fixed at the $1.43${~\AA} bonding distance of the equilateral H$_3$.
Here, we vary the angle $\alpha$ between the two sides, see Fig.~\ref{fig:molecules}(b).
We map the rectangular H$_4$ system to eight qubits with the Jordan-Wigner transformation.
The resulting $97$ Pauli terms are measured in $35$ sets of qubit-wise commuting elements.
We first find the bond distance of the square system by computing the dissociation curve with full CI.
The square H$_4$ at the $0.9${~\AA} equilibrium distance serves as a starting point to study a rectangular H$_4$ system with an angle $\alpha=40^\circ$ between the two $1.8${~\AA} fixed-length diagonals, see Fig.~\ref{fig:molecules}(c).
In general, the number of required qubits increases with the number of atoms and size of the spin orbital basis set.
By considering only the chemically active orbitals of a molecule, e.g., with an embedding scheme~\cite{Rossmannek2021, Rossmannek2023}, the number of required qubits can be reduced which helps implement VQE on noisy quantum hardware~\cite{Barkoutsos2018}.

\section{Pulse-based VQA for hydrogen-based systems\label{sec:hardware_results}}

We now study H$_2$, H$_3$, and H$_4$ as described in Sec.~\ref{sec:chem} on IBM Quantum cross-resonance-based hardware.
We compare CNOT-based and pulse-based Ans\"atze which match the qubit connectivity.
The CNOT-based Ansatz is built from the \texttt{RealAmplitude} blueprint circuit in Qiskit~\cite{Qiskit} consisting of CNOT gates sandwiched by $R_{\rm Y}(\theta)$ rotations, here decomposed to $\sqrt{X}$ and $R_Z(\theta)$ gates.

The pulse level allows an arbitrary parametrization of the controls.
In the extreme case, each sample of the arbitrary waveform generator is a control parameter to optimize.
This enables extremely short single-qubit gates without leakage but results in an optimization landscape with many parameters~\cite{Werninghaus2020}.
To make a control scheme practical, the number of parameters to optimize must be kept reasonable~\cite{Machnes2018}.
We therefore employ pulse-based Ans\"atze in which each single-qubit gate is a DRAG pulse~\cite{Motzoi2009}, indicated by $R_X(\theta)$, with an amplitude controlled by the optimizer.
This avoids the double $\sqrt{X}$ decomposition, shown in Fig.~\ref{fig:h2_circs}(a), sparing one pulse.
The duration, standard deviation, and DRAG parameter are obtained from the calibrated $X$ gate of the backend.
Entanglement is created by cross-resonance tones each implemented as a single GaussianSquare pulse, i.e., a flat-top pulse with Gaussian edges, applied to the control qubit (0) at the frequency of the target qubit (1).
The standard deviation $\sigma$ of the flanks is $64~{\rm d}t$ with each flank containing $2\sigma$.
The duration of a single sample of the arbitrary waveform generator is ${\rm d}t=0.222~{\rm ns}$.
By contrast with Ref.~\cite{Meirom2022}, both the amplitude and duration of the \texttt{CR} pulses are variational parameters and we do not introduce an echo in the CR tones to keep them short.
Crucially, the hardware only accepts pulses with a duration that is a multiple of 16 samples and an amplitude ranging from -1 to~1.
To satisfy these conditions we introduce parameter wrapping functions described in Appendix~\ref{sec:param_wrap}.
Any non-linearity resulting from changes in the pulse amplitude is dealt with by the optimizer.

\subsection{Hydrogen molecule}

\begin{figure}[tbp!]
    \centering
    \includegraphics[width=\columnwidth, clip, trim=18 15 23 0]{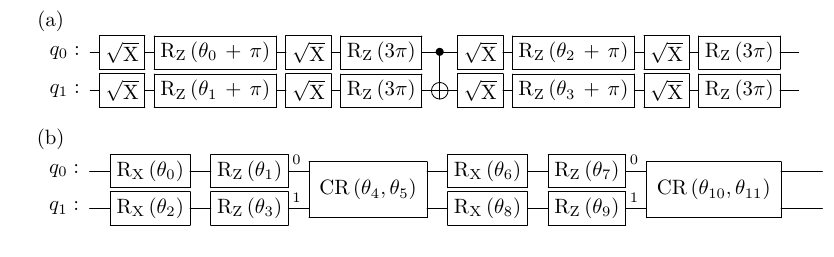}
    \caption{
    \textbf{H$\boldsymbol{_2}$ Ans\"atze.}
    In the CNOT-based variational form (a) there are four parameters in virtual-$Z$ gates.
    In the pulse-based variational form (b) there are twelve parameters.
    }
    \label{fig:h2_circs}
\end{figure}

We first run a VQE to find the ground state of H$_2$ on \emph{ibm\_lagos} on qubits 0 and 1.
We compare the circuit-based Ansatz in Fig.~\ref{fig:h2_circs}(a) to the pulse-based variational form in Fig.~\ref{fig:h2_circs}(b) in their ability to approach the exact full CI energy obtained in the chosen STO-3G molecular basis.
Simulations indicate that at least two ${\rm CR}$ pulses are needed for the pulse-based Ansatz to converge, see Appendix~\ref{sec:h2_appendix}.
Since the amplitude in the CR pulse is a real parameter we add a virtual-$Z$ gate before each CR gate to control the phase of the cross-resonance drive.

\begin{figure}[tbp!]
    \centering
    \includegraphics[width=\columnwidth]{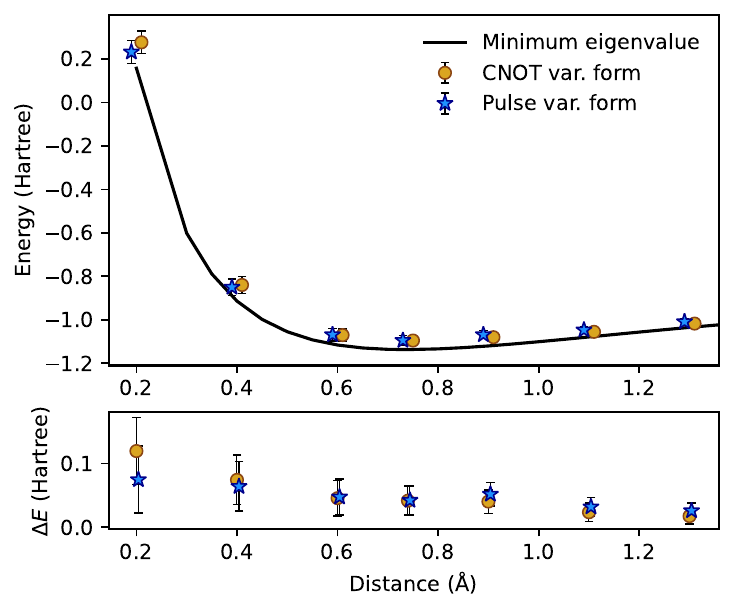}
    \caption{
    \textbf{Energy in H$\boldsymbol{_2}$.}
    The golden dots show the CNOT-based Ansatz.
    The blue stars show the pulse-based Ansatz.
    The black line is the exact energy.
    The bottom panel shows the energy difference $\Delta E$ between VQE and the exact diagonalization.
    The error bars show an upper bound on the sampling error of the estimator, see Appendix~\ref{sec:err}.
    They are increased by a factor of two and markers are slightly $x$-shifted for visibility purposes.
    }
    \label{fig:h2}
\end{figure}

\begin{figure}[htbp!]
    \centering
    \includegraphics[width=\columnwidth]{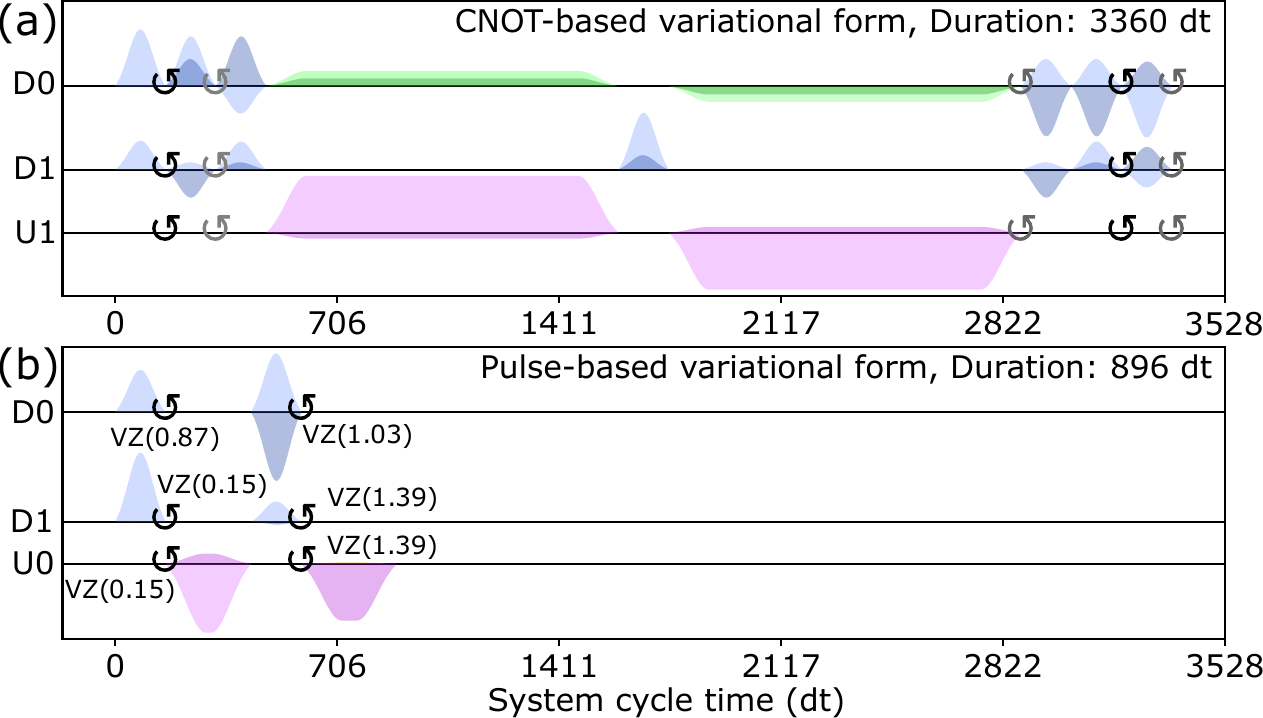}
    \caption{
    \textbf{Optimized pulse schedules for H$\boldsymbol{_2}$ at $\boldsymbol{0.2}$~\AA.}
    (a) CNOT-based and (b) pulse-based Ans\"atze.
    The circular arrows indicate zero-duration virtual-$Z$ gates (VZ).
    In the CNOT-based Ansatz the black virtual-$Z$ gates are the only parameterized instructions.
    The drive channels ${\rm D}i$ indicate single-qubit pulses on qubit $i$ and the control channels ${\rm U}j$ indicate cross-resonance tones, see details in Appendix~\ref{sec:hardware}.
    }
    \label{fig:h2_schedule}
\end{figure}

The optimization is done with COBYLA with 4096 shots per circuit.
Both the CNOT and the pulse-based Ans\"atze closely reproduce the energy, see Fig.~\ref{fig:h2}.
These results do not make use of readout error mitigation (REM) and are comparable to the non-readout error mitigated results in Ref.~\cite{Meirom2022}. 
The duration of our CNOT-based variational form, shown in Fig.~\ref{fig:h2_schedule}, is $3360~{\rm d}t$ and the duration of the pulse-based Ansatz is $928\pm114~{\rm d}t$ averaged over the considered distances $d$.
This corresponds to a difference in duration of $0.54~\mu{\rm s}$ since ${\rm d}t=0.222~{\rm ns}$.
We do not expect the pulse-based Ansatz to produce a significant gain over the CNOT-based one.
Indeed, 
(i) the noiseless CNOT-based Ansatz exactly creates the ground state of H$_2$, 
(ii) the $0.54~\mu{\rm s}$ schedule difference is small compared to the $T_1$ and $T_2$ times shown in Appendix~\ref{sec:hardware}, and 
(iii) the results are dominated by readout errors.
Overall, even in this simple example, pulse-based VQE delivers a shorter schedule than a CNOT-based VQE without affecting performance.
The next two sections show that this trend generalizes: the shorter pulse-based VQE schedules outperform their CNOT counterparts.

\subsection{Three hydrogen atoms}

\begin{figure*}[t]
    \centering
    \includegraphics[width=\textwidth,clip, trim = 10 11 15 0]{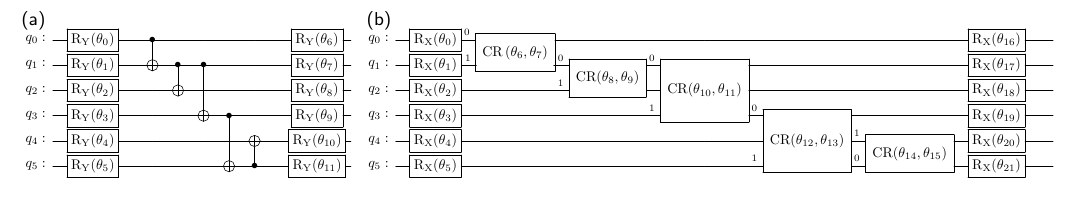}
    \caption{
    \textbf{H$\boldsymbol{_3}$ Ans\"atze.}
    (a) CNOT-based Ansatz in which a layer of parameterized $R_Y$ gates is applied before and after the CNOT gates.
    In total, the variational form has twelve parameters.
    This circuit is transpiled to the $\{\sqrt{X}, R_Z, {\rm CNOT}\}$ basis.
    (b) Pulse-based Ansatz with 22 parameters.
    Each gate corresponds to a single pulse.
    The first and second parameter in the cross-resonance gates control the duration and amplitude, respectively, of the GaussianSquare pulse.
    }
    \label{fig:h3_pulse}
\end{figure*}

\begin{figure}[htbp!]
    \centering
    \includegraphics[width=\columnwidth,clip,trim= 7 7 7 5]{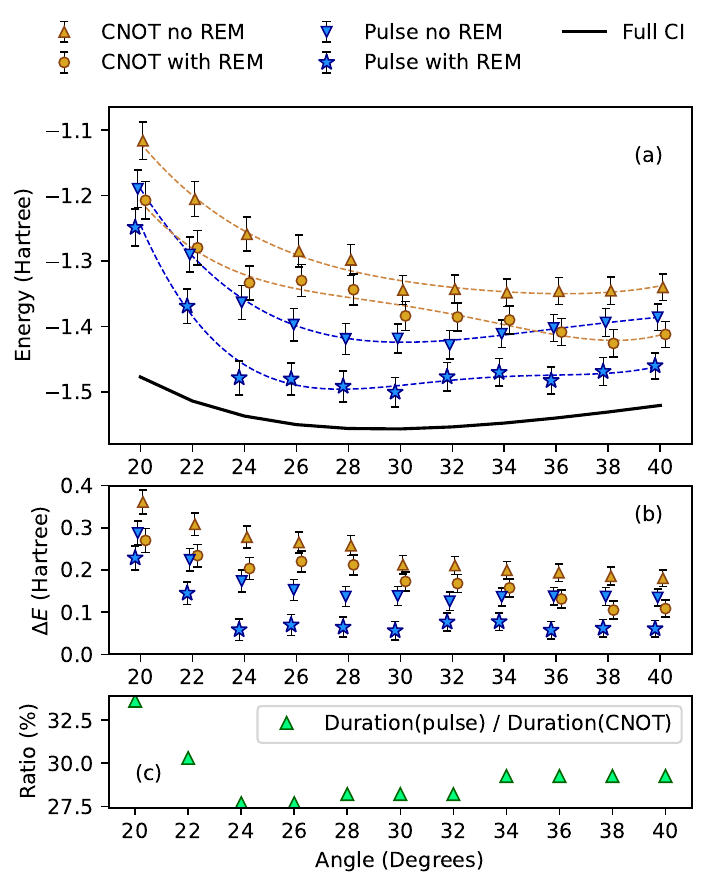}
    \caption{
    \textbf{Energy in H$\boldsymbol{_3}$.}
    (a) Sum of the electronic energy, obtained with VQE, and the repulsion energy of H$_3$ without readout error mitigation (REM) (triangles) and with REM (circles and stars).
    The CNOT-based and pulse-based Ans\"atze of Fig.~\ref{fig:h3_pulse} are labled CNOT and Pulse, respectively.
    The dashed lines are fourth order fits.
    The solid black line is the full CI solution.
    (b) Absolute difference with respect to the ideal energy.
    (c) Schedule duration of the pulse-based variational form, without REM and excluding measurement pulses, expressed as a fraction of the CNOT-based variational form which lasts $9184~{\rm d}t$, i.e. $2.04~\mu{\rm s}$.
    The error bars show an upper bound on the sampling error of the estimator, see Appendix~\ref{sec:err}.
    They are increased by a factor of two and markers are slightly $x$-shifted for visibility purposes.
    }
    \label{fig:h3_lagos}
\end{figure}

The H$_3$ system is larger than the H$_2$ molecule; it requires a total of six qubits.
We search for the ground state of H$_3$ as a function of the angle $\alpha$ on \emph{ibm\_lagos} with qubits 0, 1, 2, 3, 4, and 5.
A direct diagonalization of H$_3$ reveals a ground state with only real amplitudes at all considered angles.
We therefore compare a CNOT-based \texttt{RealAmplitude} Ansatz, shown in Fig.~\ref{fig:h3_pulse}(a), to a pulse-based one with the same structure, see Fig.~\ref{fig:h3_pulse}(b).
In both circuits the ladder of two-qubit gates matches the qubit connectivity of \emph{ibm\_lagos}, see Appendix~\ref{sec:hardware}.
For H$_3$ we focus on the depth-one CNOT-based variational form, which has twelve parameters, since deeper Ans\"atze did not improve the energy, see Appendix~\ref{sec:h3_appendix}.
The pulse-based variational form has a total of 22 optimization parameters; twelve single-qubit pulse amplitudes, five CR durations, and five CR amplitudes.
As with H$_2$, we use COBYLA with 4096 shots per circuit evaluation.

\begin{figure}[htbp!]
    \centering
    \includegraphics[width=\columnwidth]{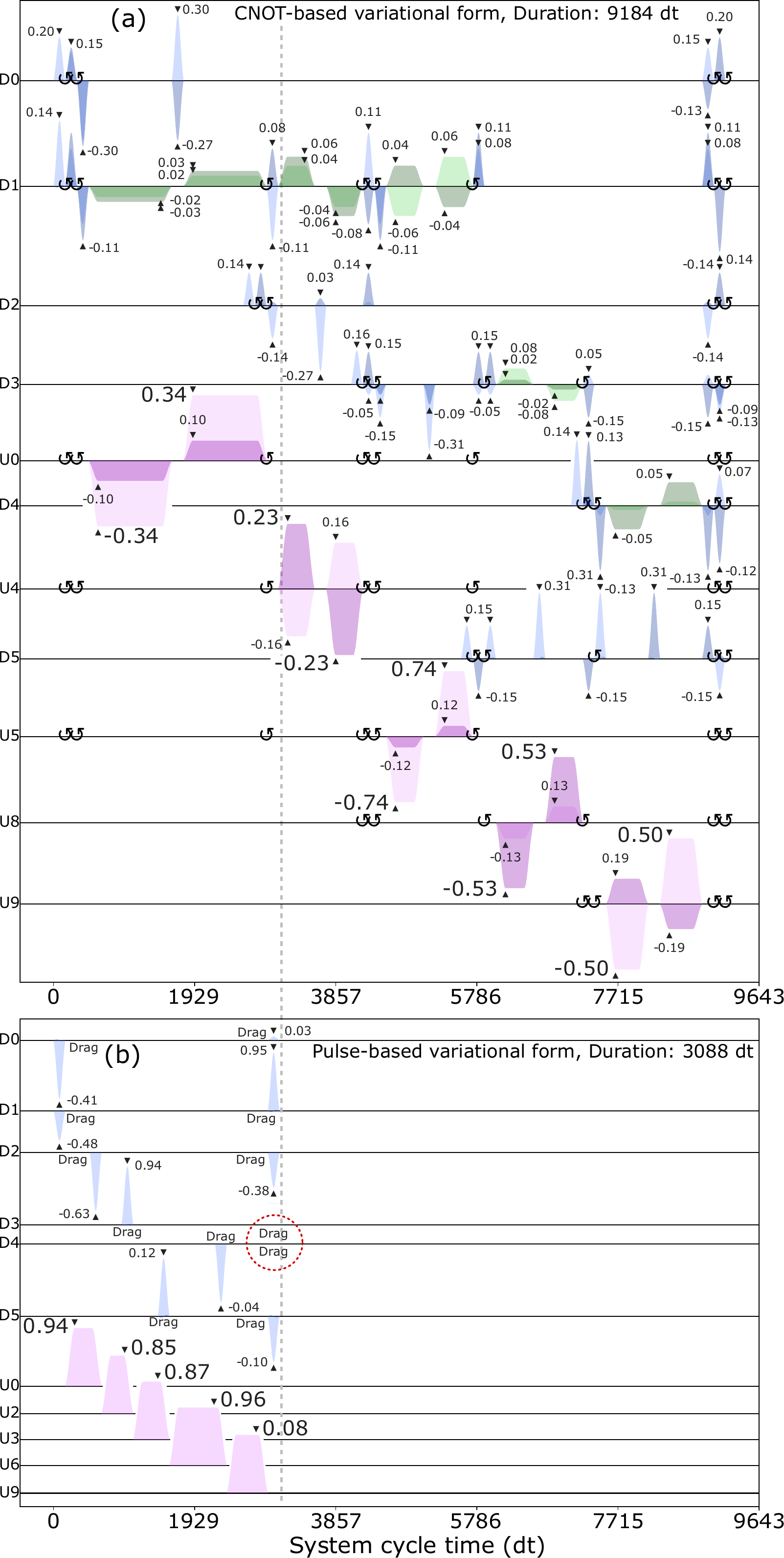}
    \caption{
    \textbf{Optimized pulse schedules for H$\boldsymbol{_3}$ at $\boldsymbol{\alpha=20^\circ}$.}
    (a) and (b) correspond to the CNOT-based and pulse-based schedules, respectively.
    The cross-resonance pulses in (b) are not necessarily applied on the same control channels as the CNOT gates in (a).
    In such a case, the role of the control and target qubits is reversed.
    The numbers indicate the amplitude of the pulse.
    The light and dark shades indicate the real and imaginary part of the pulse envelope, respectively.
    The green pulses in (a) correspond to rotary pulses.
    The dotted red circle in (b) indicates pulses in the variational form whose amplitude was set to zero by the optimizer.
    The control channel indexing is discussed in Appendix~\ref{sec:hardware}.
    The numbers above each pulse indicate the amplitude of the pulse as a fraction of the maximum output voltage of the arbitrary waveform generator.
    }
    \label{fig:lagos_h3_sched}
\end{figure}

We first find the electronic energy for the angle $\alpha=20^\circ$ starting from a random guess for the variational parameters $\boldsymbol{\theta}$.
We then find the electronic energy in increments of $2^\circ$ by initializing the optimization from the best parameters of the previous angle $\alpha$.
Next, we compute the energy by adding the repulsion energy to the VQE-computed electronic energy.
The resulting energy of the pulse-based Ansatz is $33.0\pm7.8\%$ closer, averaged over all angles, to the ideal minimum energy than the CNOT-based one, see Fig.~\ref{fig:h3_lagos}.
Note that the phases of the CR gates were not optimized.
To find the angle $\alpha_\text{min}$ that minimizes the energy of H$_3$ we fit the measured energy to fourth order polynomials shown as dashed lines in Fig.~\ref{fig:h3_lagos}.
The pulse-based and CNOT-based approaches report an $\alpha_\text{min}$ of $30.1^\circ$ and $36.4^\circ$, respectively.
Full CI yields an $\alpha_\text{min}$ of $29.3^\circ$, the pulse-based results are therefore more accurate then the CNOT-based ones.

We repeat these measurements with readout error mitigation implemented using the tensored measurement fitter in Qiskit.
Here, each CNOT-based and pulse-based VQE run is initialized with the optimal parameters found without readout error mitigation for the corresponding Ansatz and angle $\alpha$.
REM significantly reduces the errors, e.g., compare the blue stars to the blue triangles in Fig.~\ref{fig:h3_lagos}(a).
With REM the pulse-based and CNOT-based VQE report an $\alpha_\text{min}$ of $27.7^\circ$ and $38.2^\circ$, respectively.
Furthermore, with REM the pulse-based VQE has a $52\pm16\%$ lower error with respect to the full CI computation than the CNOT-based Ansatz.
Interestingly, we observe that REM lowered the absolute difference between the VQE and the full CI energy. However, it increased the deviation of $\alpha_\text{min}$ with respect to the ideal $29.3^\circ$ value as the fourth-order polynomial overfits the data.

The pulse-based Ansatz has more parameters than its CNOT counterpart which may increase its expressivity.
However, the pulse-based VQE schedule is simpler than the CNOT one, see Fig.~\ref{fig:lagos_h3_sched}.
For example, the pulse-based Ansatz only has twelve single-qubit DRAG pulses.
By contrast, the CNOT-based schedule has 45 single-qubit DRAG pulses to decompose $R_Y$ gates, implement echoes~\cite{Sheldon2016}, and fix the CNOT direction, see Appendix~\ref{sec:hardware}.
Crucially, the duration of the schedule of the optimized pulses is less than one-third of the duration of the CNOT-based schedule, see Fig.~\ref{fig:h3_lagos}(c).
After the optimization, the optimal CR pulses have an almost maximum amplitude of 1.0 and are shorter than the pulses implementing CNOT gates, compare the pulses on the control channels in Fig.~\ref{fig:lagos_h3_sched}(a) and (b).
This is consistent with mitigating decoherence.
While in general, short and intense pulses may induce leakage, it is not necessarily harmful.
Leakage can help convergence, as observed in both simulations of pulse-based VQE~\cite{Asthana2022} and gate design with optimal control~\cite{Schutjens2013}.
Our results provide further evidence of the positive impact of short and intense pulses.

\subsection{Four hydrogen atoms}

\begin{figure}[!]
    \centering
    \includegraphics[width=\columnwidth,clip,trim=20 10 15 10]{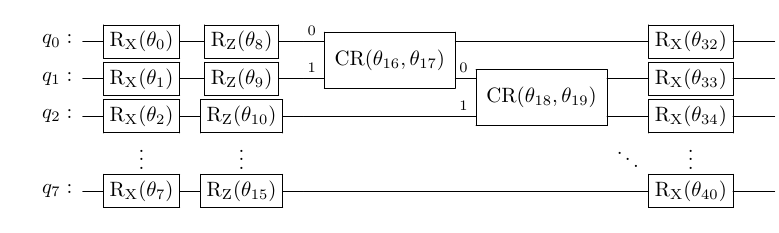}
    \caption{\textbf{H$\boldsymbol{_4}$ Pulse-based Ansatz.}
    The Ansatz has a total of 40 parameters.
    We first optimize the pulse amplitudes and durations, i.e. 32 parameters, and then the remaining 8 phases in the $R_Z(\theta_i)$ gates.
    }
    \label{fig:h4_ansatz}
\end{figure}

\begin{figure}
    \centering
    \includegraphics[width=\columnwidth, clip,trim=12 10 8 10]{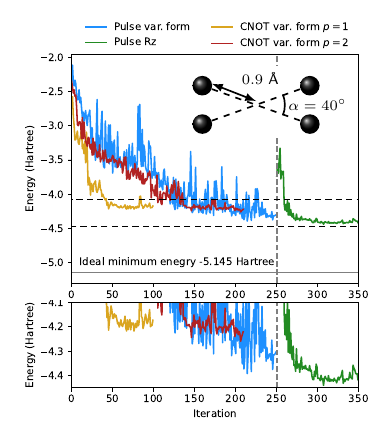}
    \caption{
    \textbf{Electronic energy of H$\boldsymbol{_4}$.}
    The gray horizontal line shows the minimum energy.
    The depth of the Ansatz is $p$.
    The bottom panel corresponds to the region between the dashed horizontal lines.
    The vertical gray line shows the point where we freeze the amplitudes and duration and optimize the phase.
    }
    \label{fig:mumbai_h4}
\end{figure}

We now evaluate the energy of the H$_4$ molecule as described in Sec.~\ref{sec:chem} at an angle of $\alpha=40^\circ$ only.
As for H$_3$, we use a \texttt{RealAmplitude} Ansatz for the CNOT-based VQE.
The pulse-based Ansatz has the same two-qubit gate structure, see Fig.~\ref{fig:h4_ansatz}.
We compare a depth-one, a depth-two CNOT-based Ansatz, and a depth one pulse-based Ansatz which have 16, 24, and 40 parameters, respectively.
For the pulse-based Ansatz we first optimize the amplitudes and durations while keeping the phases $\theta_8$ to $\theta_{15}$ at zero.
The optimization is done on \emph{ibmq\_mumbai} with COBYLA and 4096 shots per circuit.
The depth-two CNOT-based and depth-one pulse-based Ans\"atze show a similar convergence profile, see Fig.~\ref{fig:mumbai_h4}.
The pulse-based Ansatz achieves a minimum energy of $-4.39~{\rm Hartree}$ and the depth-two CNOT Ansatz achieves $-4.26~{\rm Hartree}$.

So far, we did not optimize the phases of the cross-resonance drives in the pulse-based Ansatz.
We therefore optimize the phase shifts while keeping the pulse durations and amplitudes fixed at the measured optimal values.
The initial value of the phases shifts $\theta_8,...,\theta_{15}$ is chosen at random.
These phases impact the measured energy, as seen by the decrease of the green curve in Fig.~\ref{fig:mumbai_h4}.
With the optimized phases we measure an energy of $-4.44~{\rm Hartree}$, i.e., a 1\% improvement over the pulse-based Ansatz without phases.

\begin{figure}
    \centering
    \includegraphics[width=\columnwidth]{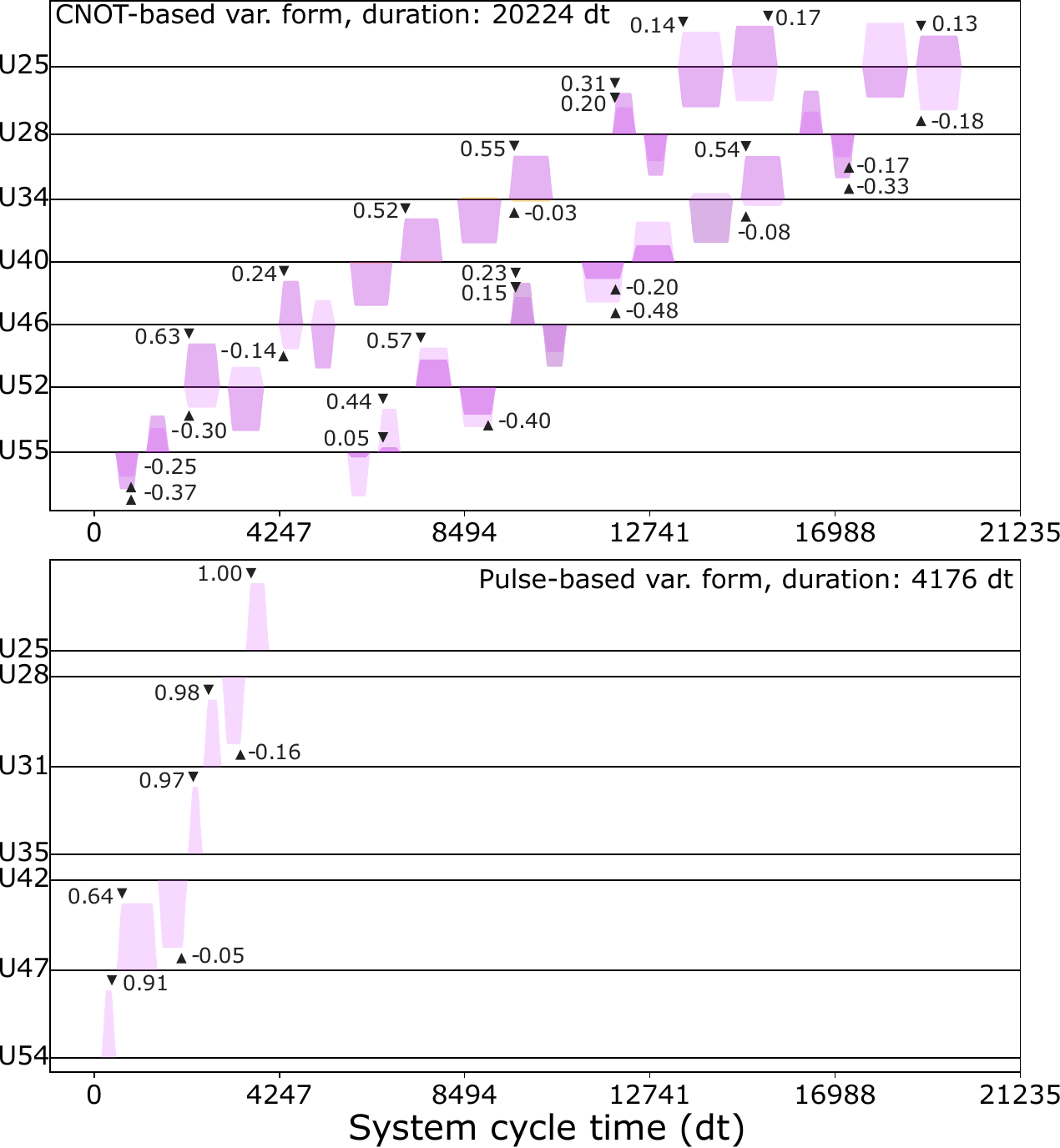}
    \caption{
    \textbf{Optimized pulse schedules for H$\boldsymbol{_4}$.} 
    The top and bottom panels show the schedules of the depth-two CNOT and pulse-based Ansatz, respectively. 
    For visualization purposes we only show the cross-resonance drives and omit the single-qubit pulses as well as the rotary tones in the CNOT gates.
    }
    \label{fig:mumbai_h4_sched}
\end{figure}

\begin{figure}
    \centering
    \includegraphics[width=\columnwidth,clip,trim=5 9 5 11]{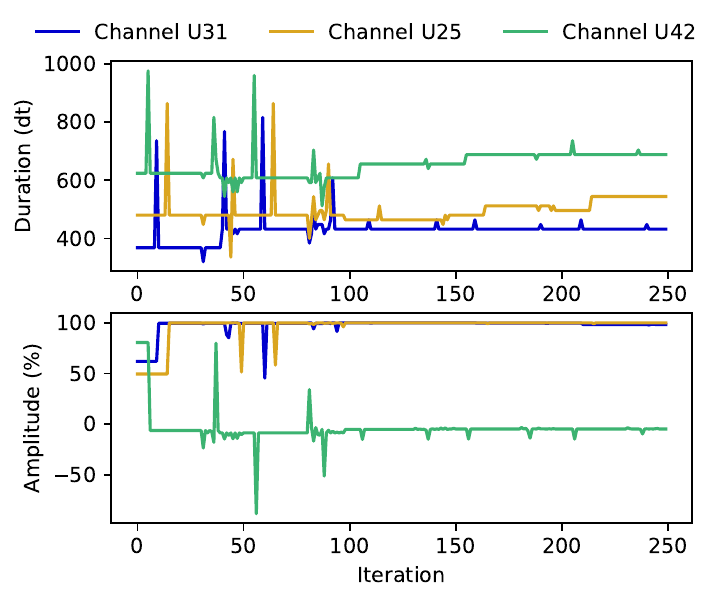}
    \caption{
    \textbf{Parameter optimization.}
    The pulse parameters of the cross-resonance pulses played on the channels \texttt{U25}, \texttt{U31}, and \texttt{U42}, shown in Fig.~\ref{fig:mumbai_h4_sched}, during the COBYLA optimization of H$_4$.
    The minimum duration of the GaussianSquare pulses is 256 samples due to the Gaussian flanks.
    }
    \label{fig:mumbai_h4_params}
\end{figure}

As for H$_3$ the optimization favours short and intense pulse schedules, see Fig.~\ref{fig:mumbai_h4_sched}.
Many of the pulses have near maximum amplitude, i.e. 1, and a short duration.
This is confirmed by inspecting the values of the pulse parameters during the optimization.
COBYLA quickly pushes up the amplitude of the cross-resonance pulses that it requires and keeps the duration small, see Fig.~\ref{fig:mumbai_h4_params}.
The best pulse-based schedule is only 20.6\% of the duration of the best depth-two CNOT-based schedule, see Fig.~\ref{fig:mumbai_h4_sched}.

\section{Error mitigation\label{sec:error_mit}}

We did not use error mitigation to focus on the gains afforded by optimizing pulse parameters.
Error mitigation may improve these results.
For example, scalable readout error mitigation, such as M3~\cite{Nation2021}, is easily applied to pulse-based VQE.
Crucially, other known error mitigation methods must be adapted to work with pulse-based VQE.
We now discuss the challenges of performing dynamical decoupling~\cite{Pokharel2018}, Pauli twirling~\cite{Wallman2016}, probabilistic error cancellation~\cite{VanDenBerg2022} and zero-noise extrapolation~\cite{Temme2017} in a pulse-based VQE.

Dynamical decoupling~\cite{viola1998dynamical,viola1998dynamical,zanardi1999symmetrizing,vitali1999using} suppresses non-Markovian errors by adding pulses in the idle regions of a schedule.
The inserted pulses mitigate the effects of decoherence~\cite{Pokharel2018,jurcevic2021demonstration} and cancel crosstalk~\cite{tripathi2022suppression,pokharel2022better} on transmon-based devices similar to those used here.
Dynamical decoupling pulses can be added in pulse-based VQE~\cite{ravi2022vaqem} in the idle regions of the Ansatz.
This may be less beneficial in pulse-based VQE than circuit VQE for three reasons.
First, the pulse-based VQE schedules are compact and have short idle regions, as exemplified by Fig.~\ref{fig:lagos_h3_sched}.
Second, idle regions change every iteration as pulse durations are optimized.
Third, the errors that dynamical decoupling suppress, e.g., crosstalk and leakage, are not always detrimental in pulse-based VQE since they can help convergence.
Nonetheless, for larger problems with longer idle delays, suppression of errors through a robust dynamical decoupling sequence could provide a noticeable performance improvement.

Pauli twirling~\cite{Wallman2016} inserts Paulis in between noisy gates, such as CNOTs and $R_{ZZ}(\theta)$ gates~\cite{Kim2021, Vazquez2022}, and commutes an inverse of each Pauli through the noisy gates.
This scheme requires the user to know the ideal operation of the noisy gate and may therefore be harder to implement in a pulse-based VQE.
For example, knowledge of the ideal cross-resonance gate could be obtained with a fit of a model to Hamiltonian tomography data.
Similarly, in probabilistic error cancellation the noise model of layers of Pauli-twirled CNOT gates is learned and corrected for in a quantum circuit by randomly inserting Pauli gates to cancel the noise on average~\cite{VanDenBerg2022}.
This is harder to implement in pulse-based VQE since the noise changes throughout the optimization as the pulse parameters are varied.

In zero-noise extrapolation an expectation value is measured several times with logically equivalent quantum circuits but with different noise levels~\cite{Temme2017}.
The noiseless expectation value is in principle recovered by extrapolating the noisy results to the zero-noise limit.
The additional noise is introduced by stretching the pulses that implement the single- and two-qubit gates~\cite{Kandala2019} or by gate folding~\cite{Lougovski2018, Giurgica2020, Stamatopoulos2020}.
Pulse stretching is hard to implement even in gate-based approaches since it requires intensive calibration and changes in pulse amplitude may induce non-linear changes in the noise.
Pulse stretching could be implemented with a large overhead in pulse-based VQE by performing tomography of the individual pulses and trying to reproduce the unitary part of the time evolution with a stretched pulse with weaker amplitude and longer duration.
By contrast, gate folding may be easier to implement, either by inserting delay instructions or by folding a pulse $P$ according to $P - [R_Z(-\pi) P R_Z(\pi) P]^n$ with $n\in\mathbb{N}$ for systems in which a negative amplitude pulse is the inverse of a positive amplitude pulse under ideal circumstances.

\section{Discussion and Conclusion\label{sec:conclusion}}

We demonstrate on hardware that VQE delivers better results when the pulse parameters such as duration and amplitude are simultaneously optimized compared to CNOT-based Ans\"atze.
For instance, a pulse-based Ansatz finds that the angle $\alpha$ that minimizes the energy of the H$_3$ system is only $2.7\%$ away from the full CI computation while a CNOT-based Ansatz measures a deviation of $24.2\%$.
Crucially, we observed that the pulse optimization favors short and intense pulses to mitigate the effects of decoherence and energy relaxation.

Our experiments are carried out on cross-resonance based hardware.
Pulse-based VQE is also applicable to other architectures such as tunable couplers~\cite{McKay2016}.
Crucially, tunable couplers support a versatile range of interactions such iSWAP, and controlled-phase generators~\cite{Ganzhorn2020} which conserve particle number.
Such exchange-like gates help reduce the circuit depth of variational Ans\"atze~\cite{Barkoutsos2018, Ganzhorn2019}.
We therefore expect that such an architecture may provide even better pulse-based Ans\"atze.
Pulse-based VQE is also applicable to other quantum computing architectures capable of pulse-shaping and variational algorithms.
For example, trapped ions and Rydberg atoms are both amenable to optimal control~\cite{Bentley2020, Jandura2022} and VQE~\cite{Nam2020, DeKeijzer2023}.
We anticipate that such systems will also benefit from the shorter schedules of pulse-based VQE as long as they are fast enough.

Running variational algorithms on hardware is time-consuming.
This makes speed a key resource for quantum computers~\cite{Wack2021}.
At the time of writing \emph{ibmq\_mumbai} reported $1800$ circuit layer operations per second (CLOPS).
Increasing the CLOPS is key to make variational algorithms scalable.
Short-duration pulse-based Ans\"atze may also help increase the CLOPS once run-time compilation and data transfer bottlenecks are removed.
Such reductions of quantum processing time are similar to restless measurements which forego qubit reset in calibration~\cite{Tornow2022} and optimal control schemes~\cite{Werninghaus2020, Werninghaus2021}.
Methods that reduce the number of shots, such as positive operator valued measures~\cite{Perez2021, Fischer2022}, are compatible with pulse-based VQE and may further reduce execution times.

The pulse-based variational forms shown here have shorter schedules but contain more parameters to optimize than gate-based ones.
This may make them more expressive but increases their optimization cost.
Future work on pulse-based VQE will need to scale-up these variational forms while keeping their parameter numbers reasonable and retaining an adequate expressiveness. We leave it to future work to investigate the expressiveness and number of parameters in pulse-based VQE in a study akin to existing research for circuit-based Ans\"atze~\cite{Sim2019}.
Methods such as ADAPT-VQE which grow the variational form one operator at a time may be modified and applied to pulse-based VQEs~\cite{Grimsley2019, Tang2021}.
Algorithms such as WAHTOR that exploit symmetries in the Hamiltonian by molecular orbital rotations could also be adapted to pulse-based Ans\"atze~\cite{Ratini2022}.
Furthermore, adapting state of the art error mitigation methods to pulse-based VQE requires more research, as discussed in Sec.~\ref{sec:error_mit}.

In summary, we showed a pulse-based Ansatz inspired by hardware constraints.
Our results demonstrated that pulse-based variational forms are a viable way to reduce schedule duration in hardware-native Ans\"atze to fight decoherence and increase the accuracy of VQE.
The quality of our results is still beyond chemical accuracy. 
As for conventional CR-based approaches, accurate results are only possible through the implementation of error mitigation schemes. 
The combination of pulse-based VQE and error mitigation will be the subject of future studies. 
These may also include the investigation of pulse-shaping methods that are closer to chemistry inspired Ans\"atze such as the unitary coupled cluster approach~\cite{Romero2018}.

\section{Acknowledgements}

This research was supported by the NCCR MARVEL, a National Centre of Competence in Research, funded by the Swiss National Science Foundation (grand number 205602). 
This research has received funding from the European Union’s Horizon 2020 research and innovation program under the Marie Sk\l{}odowska-Curie grant agreement No.~955479.
IBM, the IBM logo, and ibm.com are trademarks of International Business Machines Corp., registered in many jurisdictions worldwide. Other product and service names might be trademarks of IBM or other companies. The current list of IBM trademarks is available at \url{https://www.ibm.com/legal/copytrade}.

\appendix

\section{Numerical simulations of cross-resonance pulse-based VQE\label{sec:h2_appendix}}

We simulate pulse-based VQE with both Qiskit Aer and Qiskit Dynamics.
The pulse-based Ansatz, e.g. Fig.~\ref{fig:h3_pulse}(b), has custom \texttt{CR} instructions each encapsulating a GaussianSquare pulse as a schedule.
Before simulating the quantum circuit in Qiskit Aer we run a transpiler pass that identifies any \texttt{CR} instructions with a pulse schedule.
When such an instruction is found we attach to it a unitary matrix obtained from a Qiskit Dynamics simulation.
This simulation solves the time-evolution of a two-qubit system only, with a Hamiltonian given by Eq.~(\ref{eq:cr_eff}).
The coefficients of $\bar{H}_\text{cr}$ are measured on the hardware with Qiskit Experiments~\cite{QiskitExperiments}.

\begin{figure}[t!]
    \centering
    \includegraphics[width=\columnwidth]{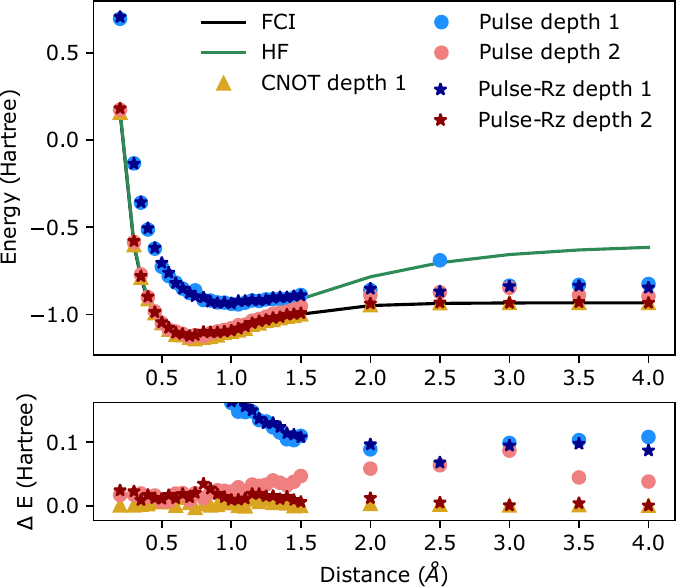}
    \caption{
    \textbf{Simulated H$\boldsymbol{_2}$ dissociation curves.}
    The total energy is plotted against the distance between the two H atoms.
    The yellow triangles correspond to the CNOT-based Ansatz.
    The circles and stars correspond to the pulse-based wave function without and with the $R_Z$ rotations, respectively.
    The black and green lines represent the full CI and Hartree-Fock energies, respectively.
    The bottom panel shows the absolute energy difference between the VQE simulations and the full CI energy.
    }
    \label{fig:h2_dissociation}
\end{figure}

We now consider the hydrogen molecule starting from the $0.74${~\AA} bond distance. 
We run a noiseless VQE for each point on the curve with COBYLA and the \emph{qasm simulator} in Qiskit with $8192$ shots.
The variational parameters $\boldsymbol{\theta}$ in each optimisation are initialized with the optimal parameters of the nearest considered bond distance, with the exception of $d=0.74${~\AA} for which random parameters were chosen.
We compare a \texttt{RealAmplitude} depth-one CNOT-based variational wave function to two pulse-based variational forms, with and without $R_Z$ rotations before the cross-resonance gates.
Moreover, each pulse-based variational form is studied at depths one and two, see Fig.~\ref{fig:h2_dissociation}.

The energy obtained from the depth-one pulse-based Ansazt shows that a single-cross resonance pulse is not sufficiently expressive, see blue markers in Fig.~\ref{fig:h2_dissociation}.
Indeed, a depth-two pulse-based Ansatz, i.e. two cross-resonance tones, is required to get energies close to the full CI dissociation curve, see red circles in Fig \ref{fig:h2_dissociation}.
Furthermore, adding $R_Z$ gates to control the phase of the cross-resonance tones allows us to recover almost all of the system's correlation energy, see the red stars. 
The CNOT-based Ansatz directly engineers the ground state of H$_2$.

\section{Pulse parameter wrapping\label{sec:param_wrap}}

\begin{figure}[tb!]
    \centering
    \includegraphics[width=\columnwidth]{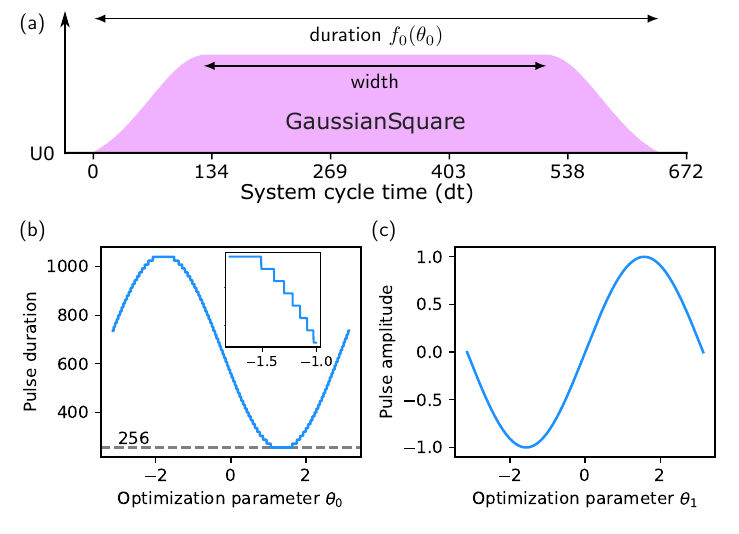}
    \caption{\textbf{Example of wrapper functions.}
    (a) GaussianSquare pulse with duration $f_0(\theta_0)$ and amplitude $f_1(\theta_1)$.
    (b) Duration wrapper that relates the optimization parameter $\theta_0$ to the pulse duration through a sinusoidal function vertically discretized to multiples of 16. 
    The minimum duration is set to 256 samples which corresponds to a width of 0.
    (c) Amplitude wrapper that relates the optimization parameter $\theta_1$ to the amplitude of the pulse through a sinusoidal function.
    }
    \label{fig:wrappers}
\end{figure}

Qiskit Pulse allows users to manipulate quantum computers at the level of pulses by specifying schedules of pulses~\cite{Mckay2018, Alexander2020}.
The pulses must satisfy hardware imposed requirements.
First, the complex-valued pulse amplitude is expressed as a fraction of the maximum output voltage of the arbitrary waveform generator (AWG) and must therefore be restricted to the interval $[-1, 1]$.
Second, the duration of a pulse, expressed in the number of AWG samples, must be a multiple of 16 to be loaded in the AWG memory.
Third, the duration must be kept large enough to prevent the width of the flat-top from being negative, see Fig.~\ref{fig:wrappers}(a). 
To impose these constraints we introduce the concept of a parameter wrapper.
In a pulse-based variational form the optimizer optimizes the parameters $\theta_i$.
However, the parameters used to construct the pulses are the output of functions that wrap $\theta_i$, i.e., $f_i(\theta_i)$.
The amplitude is restricted to the interval $[-1, 1]$ by a sinusoidal function, see Fig.~\ref{fig:wrappers}(c).
The duration is restricted to the interval [256, 1040] samples by a sinusoidal function whose codomain is restricted to multiples of 16.
The lower-bound ensures that the width does not become negative and the upper bound,  loosely chosen based on the strength of $\omega_{ZX}$, prevents the pulse from becoming too long.

\section{Statistical error bound~\label{sec:err}}

Here we derive an upper bound for the worst-case statistical sampling error $\epsilon_\text{max}$ of the energy expectation values shown as error bars in Figs.~\ref{fig:h2} and~\ref{fig:h3_lagos}. 
We partition the Hamiltonian $H$ of each hydrogen system into $M$ groups of mutually qubit-wise commuting Pauli terms
\begin{equation}
\label{eqn:grouped_Hamiltonian}
H=\sum_{i=1}^M \sum_{j=1}^{m_i} c_{ij} P_{ij}.
\end{equation}
Here, $m_i$ is the number of Pauli terms in group $i$, $P_{ij}$ is the $j$-th Pauli in group $i$, and $c_{ij}$ are coefficients.
The number of shots used to estimate the Paulis in group $i$ is $n_i=4096~\forall~i=1,...,M$. 
Since every group is estimated independently, the total variance $\epsilon^2$ of the estimator of $\langle H \rangle$ is the sum of the standard errors of every group. 
We can thus compute the following bound 
\begin{align}
\label{eqn:bound_derivation}
\epsilon^2 &= \sum\limits_{i=1}^M \frac{1}{n_i}\text{Var}\left[ \sum_{j=1}^{m_i} c_{ij} P_{ij}  \right] \\
    &= \sum\limits_{i=1}^M \frac{1}{n_i} \sum_{j, j'=1}^{m_i} c_{ij} c_{ij'} \text{Cov}\left[ P_{ij},  P_{ij'} \right] \\
    &\leq \sum\limits_{i=1}^M \frac{1}{n_i} \sum_{j, j'=1}^{m_i} |c_{ij} c_{ij'}| \sqrt{\text{Var}\left[P_{ij} \right] \text{Var}\left[ P_{ij'} \right] } \\
    &\leq \sum\limits_{i=1}^M \frac{1}{n_i} \sum_{j, j'=1}^{m_i} |c_{ij} c_{ij'}|.
\end{align}
We used the Cauchy-Schwarz inequality and the fact that $\text{Var}\left[P_{i,j} \right] \leq 1$. 
The standard error of the estimator for any state is thus upper bounded by
\begin{align}
\label{eqn:error_bound}
\epsilon_\text{max} = \sqrt{\sum\limits_{i=1}^M \frac{1}{n_i} \sum_{j, j'=1}^{m_i} |c_{i,j} c_{i,j'}|}.
\end{align}

\section{Additional data for the three hydrogen atoms\label{sec:h3_appendix}}

\begin{figure}[t!]
    \centering
    \includegraphics[width=\columnwidth, clip, trim=8 8 8 5]{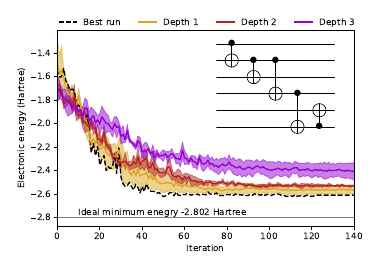}
    \caption{
    \textbf{VQE on H$\boldsymbol{_3}$ with different depths}. 
    The inset shows the structure of the entangler that was repeated $p$ times with $p+1$ layers of parameterized $R_Y(\theta)$ gates.
    The bond distance is $1.43${~\AA} and the angle is $40^\circ$.
    The solid lines and shaded areas show the mean and standard deviation, respectively, of three runs.
    The dashed black line shows the best run which was a depth-one run.
    }
    \label{fig:h3_cx_only}
\end{figure}

Here, we present additional data on the H$_3$ system.
We investigate the effect of the depth of the CNOT-based variational form on the measured energy. 
The data, made of nine hardware runs, three at each depth $p\in\{1,2,3\}$, are acquired on \emph{ibm\_lagos}.
Each run has a different initial point chosen uniformly in the interval $[0,\pi]^{6(p+1)}$.
Under these settings we observe that the best results are obtained with a depth-one ansatz, see Fig.~\ref{fig:h3_cx_only}.
Ans\"atze with depth $p>1$ did not result in a lower energy than $p=1$.
This may be due to the COBYLA optimizer getting stuck in local minima or due to the added noise of the deeper circuits.
Error mitigation methods may help to overcome the added noise while methods that progressively build-up the ansatz may help tackle local minima~\cite{Tang2021}.
The jobs took a total of $71\pm 9$, $102\pm 2$, and $99\pm 0$ minutes of classical and quantum compute time, for depths one, two, and three, respectively, as reported by \emph{ibm\_lagos}.
Each circuit was executed with 4096 shots.

\section{Hardware\label{sec:hardware}}

\begin{table*}[htb!]
    \centering \footnotesize
    \begin{tabular}{l r r r r r r | l r r r r r r} \hline\hline
        \multicolumn{6}{l}{\emph{ibm\_lagos}} & & \multicolumn{6}{l}{\emph{ibmq\_mumbai}} \\
        Qubits & $\omega_{ZX}$ & $\omega_{ZY}$ & $\omega_{ZZ}$ & $\omega_{IX}$ & $\omega_{IY}$ & $\omega_{IZ}$ & Qubits & $\omega_{ZX}$ & $\omega_{ZY}$ & $\omega_{ZZ}$ & $\omega_{IX}$ & $\omega_{IY}$ & $\omega_{IZ}$ \\\hline
        (0, 1) & $872(2)$ & $705(2)$ & $-6(2)$ & $-839(2)$ & $-584(2)$ & $14(2)$ & %
        (12, 13) & $-930(1)$ & $-637(2)$ & $93(1)$ & $-214(1)$ & $-143(2)$ & $6(1)$ \\
        (1, 2) & $-1803(2)$ & $-1056(4)$ & $7(3)$ & $-490(2)$ & $-244(4)$ & $-49(3)$ & %
        (13, 14) & $1481(1)$ & $294(4)$ & $23(2)$ & $-765(1)$ & $-170(4)$ & $-74(2)$ \\
        (1, 3) & $-2430(3)$ & $858(10)$ & $-88(8)$ & $4168(3)$ & $-1459(10)$ & $-124(8)$ & %
        (14, 16) & $520(1)$ & $-450(2)$ & $-53(2)$ & $-589(1)$ & $533(2)$ & $-140(2)$ \\
        (3, 5) & $194(1)$ & $641(1)$ & $-48(1)$ & $47(1)$ & $70(1)$ & $39(1)$ & %
        (16, 19) & $768(1)$ & $-295(2)$ & $53(1)$ & $-341(1)$ & $131(2)$ & $-19(1)$ \\
        (5, 4) & $-383(4)$ & $1660(1)$ & $-99(2)$ & $-89(4)$ & $364(1)$ & $83(2)$ & %
        (19, 22) & $-1078(1)$ & $734(2)$ & $137(2)$ & $-83(1)$ & $68(2)$ & $-64(2)$ \\
        & & & & & & & (22, 25) & $-1110(1)$ & $-509(2)$ & $4(2)$ & $396(1)$ & $181(2)$ & $-40(2)$ \\
        & & & & & & & (25, 26) & $631(2)$ & $467(3)$ & $1(2)$ & $1228(2)$ & $917(3)$ & $-20(2)$ \\

        \hline\hline
    \end{tabular}
    \caption{
    Strength of the terms in the effective cross-resonance model for \emph{ibm\_mumbai} and \emph{ibmq\_mumbai}.
    All numbers are in ${\rm kHz}$ and were measured with a single Gaussian square pulse with unit amplitude.
    }
    \label{tab:mumbai_ham_tomo}
\end{table*}

\begin{figure}[htbp!]
    \centering
    \includegraphics[width=0.95\columnwidth]{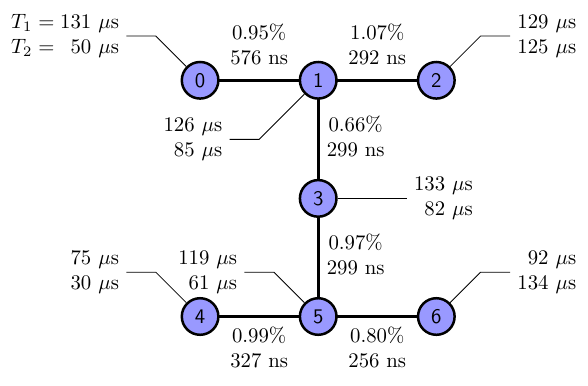}
    \caption{\textbf{Coupling map of \emph{ibm\_lagos}.}
    The numbers attached to each qubit represent the $T_1$ and $T_2$ times as reported by the backend.
    The top and bottom numbers of each edge indicate the error of the CNOT gate and its duration, respectively.
    }
    \label{fig:lagos_cmap}
\end{figure}

We now describe the hardware on which the data were gathered.
The H$_2$ molecule and the H$_3$ system were both run on the seven qubit IBM Quantum device $\emph{ibm\_lagos}$ whose coupling map is shown in Fig.~\ref{fig:lagos_cmap}(a).
Here, the CNOT gates are implemented with echoed cross-resonance pulses.
Since calibration is time-consuming the backends only calibrate one CNOT gate for each pair of coupled qubits $(i, j)$.
This CNOT gate is referred to as \emph{hardware-native}. 
The CNOT gate in the reverse direction $(j, i)$ is implemented with additional single-qubit pulses and the hardware-native CNOT gate.
The CNOT-based and pulse-based variational forms in the main text sometimes differ in the control channels on which they apply CR pulses despite the fact that the CNOT and \texttt{CR} gates are applied on the same qubit pair $(i, j)$.
This is because the desired CNOT gate may not be hardware native.
For convenience we summarize the configuration of the control channels as $(i,j)$:~${\rm U}k$ where ${\rm U}k$ is the control channel which drives qubit $i$, the control, at the frequency of qubit $j$, the target.
On \emph{ibm\_lagos} the control channel configuration is $(0, 1)$:~${\rm U}0$, $(1, 0)$:~${\rm U}1$, $(1, 2)$:~${\rm U}2$, $(2, 1)$:~${\rm U}4$, $(1, 3)$:~${\rm U}3$, $(3, 1)$:~${\rm U}5$, $(3, 5)$:~${\rm U}6$, $(5, 3)$:~${\rm U}8$, $(4, 5)$:~${\rm U}7$, and $(5, 4)$:~${\rm U}9$.

\begin{figure}[htbp!]
    \centering
    \includegraphics[width=\columnwidth]{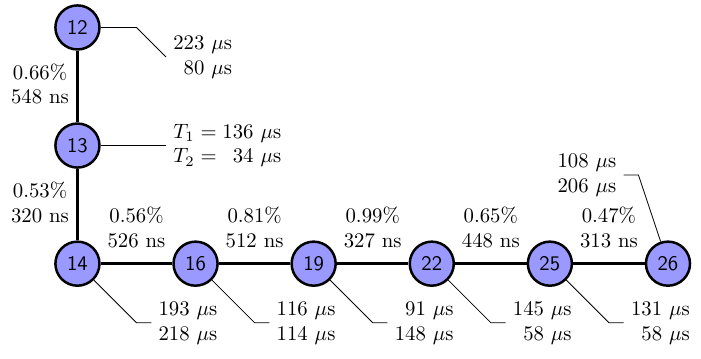}
    \caption{\textbf{Qubits of \emph{ibmq\_mumbai} to run H$_4$.}
    The numbers attached to each qubit represent the $T_1$ and $T_2$ times as reported by the backend.
    The top and bottom numbers of each edge indicate the error of the CNOT gate and its duration, respectively.
    }
    \label{fig:mumbai_cmap}
\end{figure}

The H$_4$ system was run on the linearly coupled qubits 12, 13, 14, 16, 19, 22, 25, and 26 of the 27 qubit system \emph{ibmq\_mumbai}.
The properties of these qubits are summarized in Fig.~\ref{fig:mumbai_cmap}.

\newpage

\bibliography{references}

\end{document}